\documentclass[10pt]{article}

\usepackage{amsmath}

\newcommand{\tmc}{T_\mathrm{mc}}

\usepackage{graphicx}
\begin{document}

\title{Optimized Monte Carlo Method for glasses}
\author{L.~A.~Fern\'andez$^{1,3,}$\footnote{{\tt
laf@lattice.fis.ucm.es},\ $^{\dagger}${\tt victor@lattice.fis.ucm.es},\  $^{\ddagger}${\tt
paolo.verrocchio@unitn.it}}, V.~Mart\'\i{}n-Mayor$^{1,3,\dagger}$ and
P.~Verrocchio$^{2,3,4\ddagger}$\\ {\small $^1$ Dep. de F\'\i{}sica Te\'orica I,
U. Complutense, 28040 Madrid, Spain.}\\
{\small $^2$ Dip. di Fisica, U. di Trento, 38050 Povo, Trento, Italy.}\\ 
{\small $^3$Instituto de Biocomputaci\'on y
F\'{\i}sica de Sistemas Complejos (BIFI). Zaragoza, Spain.}\\
{\small $^4$Research center Soft INFM-CNR c/o U. di Trento 38050 Povo, Trento, Italy.}}

\date{\today} \maketitle
\begin{abstract}
A new Monte Carlo algorithm is introduced for the simulation of
supercooled liquids and glass formers, and tested in two model
glasses. The algorithm is shown to thermalize well below the Mode
Coupling temperature and to outperform other optimized Monte Carlo
methods.  Using the algorithm, we obtain finite size effects in the
specific heat.  This effect points to the existence of a large
correlation length measurable in equal time correlation functions.
\end{abstract}


\newpage
\section*{\normalsize 1. Introduction}

The lack of structural or thermodynamic changes at the glass
transition~\cite{DeBenedetti97} is a major problem for its
investigation. The only standard feature, as compared with
second-order phase transitions~\cite{FSSBOOK}, is the
dramatic dynamical slowing down at the critical temperature.  A fairly
standard mechanism for slow dynamics in an homogeneous system at
finite temperature is the divergence of a correlation length ({\em
critical slowing down}~\cite{FSSBOOK}). Slowness arises from the
need of configurational changes to propagate over increasingly large
regions (the critical origin of the Mode Coupling
singularity\cite{goetze:1992} has been recently
recognized\cite{biroli:2004}).  It has been recently
proposed~\cite{FERNANDEZ06} to study this growing lengthscale in
glassformers through the finite size behaviour of small
systems~\cite{FSSBOOK}.  Note that experiments in
films and nanopores\cite{FSS-REVIEW1,FSS-REVIEW2} show that the
glass transition changes in samples with one or more dimensions of
nanometric scale. In particular, the specific heat is most sensitive
to the size of the confining pore when temperatures are close to the
glass transition\cite{FSS-PORES}.

Numerical simulations are an important tool for the study of the glass
transition.  Their worse drawback is the shortness of the times that
may be simulated in today computers (roughly speaking,
microseconds). As a consequence, the computer model goes out of
equilibrium by the Mode-Coupling temperature, $\tmc$, rather than the
actual glass temperature, $T_\mathrm{c}$.  To approach $T_\mathrm{c}$,
one may resort to optimized Monte Carlo (MC)
methods~\cite{Grigera01,dePablo,FERNANDEZ06}, namely methods
implementing unphysical dynamical rules that strongly reduce the
equilibration times. When thermalization is achieved, optimized MC
allow to study {\em equilibrium} mean values and their temperature (or
pressure) derivatives, although the purely dynamic features of these
methods are interesting on their own right~\cite{FERNANDEZ06}.

Here, we give the first full description of the local swap
algorithm~\cite{FERNANDEZ06}. We compare the performance of the local
swap dynamics with the standard MC and with the microcanonical
algorithm~\cite{dePablo}.  We conclude that local swap yield
equilibrium data at temperatures where the microcanonical algorithm no
longer thermalizes. Finally, we address the fishy issue of estimating
the specific heat in a {\em metastable} liquid state. We give here
details on the strategy followed in~\cite{FERNANDEZ06}, where tiny but
clearly measurable finite-size effects were observed in the specific
heat.

\section*{\normalsize 2. Models and observables}

We consider two similar models of fragile glass formers, namely binary
mixtures of soft spheres. The first model, extensively studied in
Ref.~\cite{FERNANDEZ06}, is a 50\% mixture of particles interacting
through the pair potential
$V_{\alpha\beta}(r)=\epsilon[(\sigma_\alpha+\sigma_\beta)/r]^{12} +
C_{\alpha\beta}$, where $\alpha,\beta=A,B$, with a cutoff at
$r_\mathrm{c}=\sqrt{3}\sigma_0$. The choice $\sigma_A=1.2\sigma_B$
hampers crystallization, as compared with the $\sigma_A=\sigma_B$
model. We impose
$(2\sigma_A)^3+2(\sigma_A+\sigma_B)^3+(2\sigma_B)^3=4\sigma_0^3$ where
$\sigma_0$ is the unit length.  Constants $C_{\alpha \beta}$ are
chosen to ensure continuity at $r_\mathrm{c}$. The simulations are at
constant volume, with particle density fixed to $\sigma_0^{-3}\,$ and
temperatures in the range $[0.897 T_\mathrm{mc}, 10.792
T_\mathrm{mc}]$. We use periodic boundary conditions on a box of size
$L$ in systems with $N=512,1024,2048$ and $4096$ particles.  For argon
parameters, $\sigma_0=3.4$\AA, $\epsilon/k_B=120$K and
$T_\mathrm{mc}=26.4$K.

In the second model~\cite{CARRUZZO04,dePablo} the choice
$\sigma_B=1.4\sigma_A$ is made. Naming
$x=r/(\sigma_\alpha+\sigma_\beta)$, the pair potential in units
$\epsilon=1$ is $V(x)=x^{-12}+x-13/12^{12/13}$ if $x<12^{1/13}$ and
zero otherwise (thus, the cut-off distance depends on the type of
interacting species). We study density $1.08\sigma_0^{-3}\,$, as in
Refs.~\cite{CARRUZZO04,dePablo}.

Since the potential energy per particle, $e$ shows the slowest
excitations~\cite{FERNANDEZ06,CARRUZZO04}, we shall focus here only in
this observable, the internal energy being $\frac32 k_\mathrm{B} T
+\langle e\rangle$ (for other quantities, see
Ref.~\cite{FERNANDEZ06}).  The constant-volume specific heat, $C_v$,
is:
\begin{equation}
e= \frac{1}{2N} \sum_{j,k\neq j} V({\vec r_k -\vec r_j})\,,\quad
C_v=\frac32 + \frac{N}{T^2}\left[\langle e^2\rangle -\langle
e\rangle^2\right]\,\quad(\mathrm{units}\
\epsilon=k_\mathrm{B}=1).\label{DEFINITIONS}
\end{equation}

\section*{\normalsize 3. The local swap Monte Carlo algorithm}
\begin{figure}
\begin{center}
  \includegraphics[angle=90,width=0.6\columnwidth]{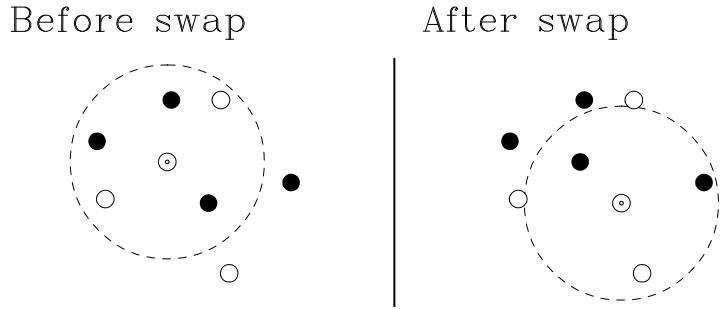}
  \caption{The local swap move. The A particles are depicted by full
symbols, while B particles are open symbols. The (say) B particle
picked randomly (signalled by a central mark) may be swapped with any
one of the three A particles inside the sphere shown in the left part
of the plot. After the swap (right), only two B particles could be
exchanged with the picked B particle.\protect{\label{FIG1}}}
\end{center}
\end{figure}

The Grigera-Parisi swap algorithm\cite{Grigera01} consists in picking
randomly a pair of particles of distinct type, $A$ and $B$, try to
exchange their positions, and accepting this move with probability
$\mathrm{min}\{1,\mathrm{e}^{-\Delta E/T}\}$ ($\Delta E$ is the total
potential energy change produced by the swap).  If combined with
standard MC, it is very effective in reducing the equilibration
time. Nevertheless, there is a caveat: the acceptance of the swap move
is very small, and it significantly decreases when the number of
particles increases. Indeed, the closer the swaped particles are, the
larger the swap acceptance becomes (this is a huge effect). With large
systems, it is higly improbable to pick for the swap neighboring
particles. To cure this problem, we have proposed the local swap.

In the local swap, the elementary MC step is either (with probability
$p$) a single-particle displacement attempt or (with probability
$1-p$) an attempt to {\em swap} particles. Therefore, for $p=1$ the
algorithm reduces to standard MC. From here on we call local swap to
the algorithm with $p=0.5$. The time unit $t_0$ is $N/p$ elementary
steps.  Our swap consists in picking a particle at random (the {\em
picked} particle) and trying to interchange its position with that of
a particle of opposite type (the {\em swapped} particle), chosen at
random among those at distance smaller than $0.6 r_\mathrm{c}$ (see
Fig.~\ref{FIG1}).  Yet, there is an intrinsic lack of
symmetry. Indeed, let $N_\mathrm{old}$ be the number of swappable
particles around the picked particle in its original position, and
$N_\mathrm{new}$ the number of swappable particles for the picked
particle in its final position. The probability of choosing the swaped
particle is $1/N_\mathrm{old}$ in the original configuration, while it
would be $1/N_\mathrm{new}$ in the final configuration.  Detailed
balance (see e.g.~\cite{FSSBOOK}) holds only if this asymmetry is
incorporated in the probability of accepting the swap:
\begin{equation}
p_\mathrm{accept\
swap}=\mathrm{min}\{1,\frac{N_\mathrm{old}}{N_\mathrm{new}}\mathrm{e}^{-\Delta
E/T}\}\,.
\end{equation}

\begin{figure}
\begin{center}
  \includegraphics[angle=270,width=0.9\columnwidth]{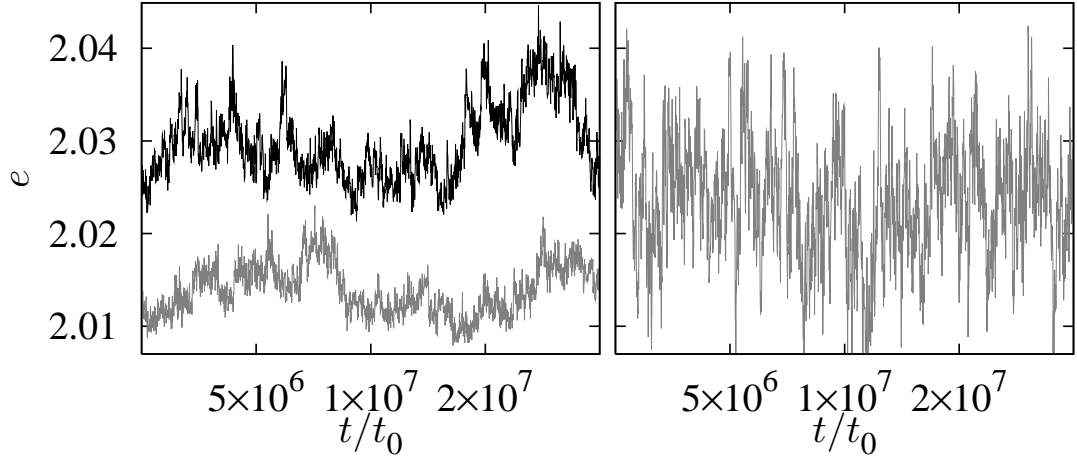}
  \caption{Comparison of the performance of the local swap (right) and
standard Monte Carlo (left), as shown by the Monte Carlo history of
the potential energy per particle (data points are the average of
$10^4 t_0$ succesive steps), for the $\sigma_A/\sigma_B=1.2$ model at
$T=0.897\tmc$ and 1024 particles. The two standard Monte Carlo runs
had as starting points thermalized configurations obtained with local
swap. In our time window, the two standard simulations do not explore
the same energy range. Instead the single local swap simulation
explores the full energy range.\protect{\label{FIG2}}}
\end{center}
\end{figure}

The acceptance of the local swap is independent of the number of
particles, and larger by a factor of 10 than for the original swap
algorithm~\cite{Grigera01}, already for $N=1024$.  For the model
$\sigma_A/\sigma_B=1.2$, the acceptance varies from 0.74\% at
0.9$T_\mathrm{mc}$ up to 6\% at $2T_\mathrm{mc}$. For the
$\sigma_A/\sigma_B=1.4$ is much smaller, actually of the order of
$8\times 10^{-6}$ at $T=0.83\tmc$, as could be guessed by the
disparity in particle diameters.

\begin{figure}
\begin{center}
\includegraphics[angle=270,width=0.9\columnwidth]{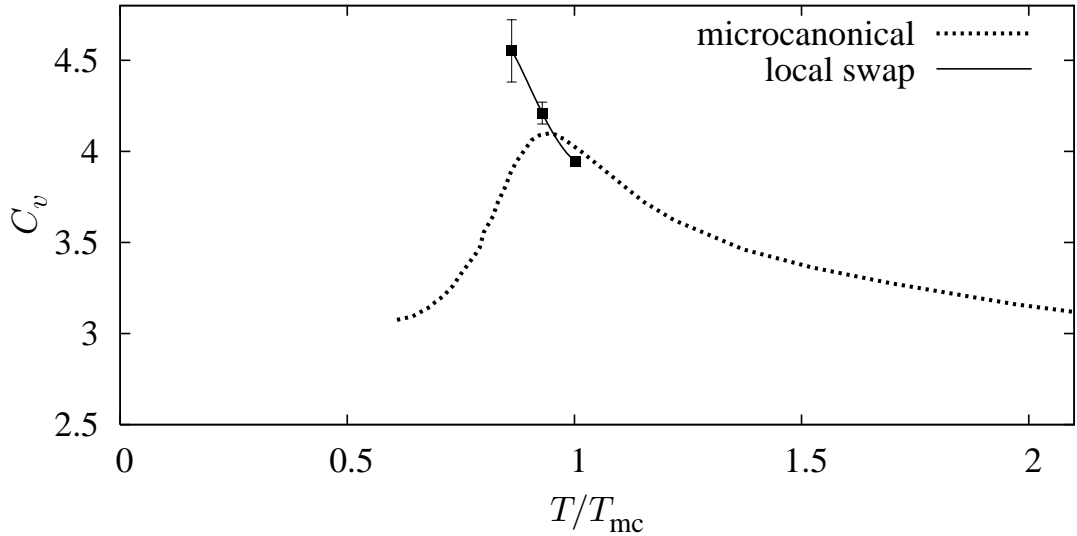}
\caption{Specific heat of the $\sigma_A/\sigma_B=1.4$ model with 128
particles, as a function of temperature, as obtained with local swap
($10^9 t_0$ steps) and with the microcanonical method~\cite{dePablo}
(error estimates were not provided in Ref.~\cite{dePablo}).  Local
swap produces a significantly larger estimate of the specific heat at
our lowest simulated temperatures, signalling better sampling of
configuration space and indicating that the microcanonical method is
unable to thermalize at such low temperatures.\protect{\label{FIG3}}}
\end{center}
\end{figure}

The performance of the local swap below $\tmc$ is far superior to the
standard MC (Fig.\ref{FIG2}) and to the microcanonical method
of Ref.~\cite{dePablo} (Fig.\ref{FIG3}). Furthermore,
for both the $\sigma_A/\sigma_B=1.2$ and the $\sigma_A/\sigma_B=1.4$
models, local swap finds a crystallization phase transition, to highly
disordered crystals on the bcc family, not reported on previous
studies~\cite{CARRUZZO04,dePablo}.

\section*{\normalsize 4. Finite size effects}

\begin{figure}
\begin{center}
\includegraphics[angle=270,width=0.9\columnwidth]{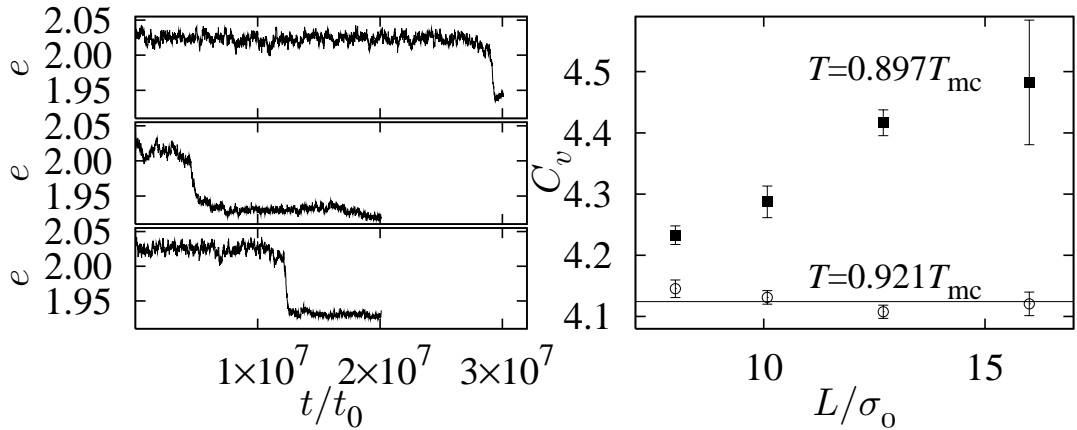}
\caption{(Left) Examples from the 100 generated (local swap) Monte
Carlo histories for the potential energy density of the
$\sigma_A/\sigma_B=1.2$ model with 1024 particles at
$T=0.897\tmc$. One may clearly distinguish the metastable liquid from
the crystallizing system, that starts as a sharp energy drop in all
three runs.  (Right) Specific heat for the model
$\sigma_A/\sigma_B=1.2$ for sistems with $512,1024,2048$ and $4096$
particles, at $T=0.921\tmc$ (empty symbols) and $T=0.897\tmc$ (full
symbols), versus the size of the simulation box (the horizontal line
is a fit of the $T=0.921\tmc$ data to a constant value, with
$\chi^2/\mathrm{d.o.f.}=4.89/3$).  At $T=0.897\tmc$, the specific heat
increases with system size, while the energy density (not shown), is
independent of the number of particles, within our accuracy.
\protect{\label{FIG4}}}
\end{center}
\end{figure}

Even if not previously known, the liquid state is metastable in our
soft-sphere models. Thus, the question arises of how to study the
thermodynamic properties of a metastable phase.  The underlying
assumption is that the equilibration time for the metastable liquid
phase is much smaller than the crystallization time. Our strategy has
been to run several MC runs (up to 400 at the lowest
temperatures~\cite{FERNANDEZ06}). On each run, the equilibrated
metastable liquid is neatly separated from the crystallizing system
(see Fig.~\ref{FIG4}-left). In the analysis we only consider histories
whose metastable liquid part was selfconsistenly found to be longer
than 100 exponential autocorrelation times (to avoid bias, we also
discarded some 20 autocorrelation times {\em before} crystallization).
Note that the central MC history in Fig.~\ref{FIG4}-left does
not meet this criterium (at this low temperature the autocorrelation
time was $10^5 t_0$). Then, one uses Eq.(\ref{DEFINITIONS}), with the
metastable liquid part of the acceptable histories to obtain the
results in Figure~\ref{FIG4}-right (data from Ref.~\cite{FERNANDEZ06}).

\section*{\normalsize 5. Conclusions}
We have given the first full description of the local swap
algorithm~\cite{FERNANDEZ06}. We compare the performance of the local
swap dynamics with the standard MC and with the
microcanonical algorithm~\cite{dePablo}, concluding that local swap
yield equilibrium data at temperatures where the microcanonical
algorithm no longer thermalizes. Furthermore, local swaps finds
crystal states not reported in previous work~\cite{dePablo}.  We have
shown how to estimate the specific heat in a {\em metastable} liquid
state, finding at low temperatures tiny but clearly measurable
finite-size effects~\cite{FERNANDEZ06}.

\section*{\normalsize Acknowledgments}
We thank G. Biroli, G. Parisi and R. de Nalda for discussions.  We
were partly supported by BSCH---UCM and by MEC (Spain), through
contracts BFM2003-08532, FIS2004-05073 and FPA2004-02602.  Simulations
were carried out at BIFI.

\bibliography{molveno}

\end{document}